\documentstyle[11pt,newpasp,twoside]{article}
\def\edcomment#1{\iffalse\marginpar{\raggedright\sl#1\/}\else\relax\fi}
\reversemarginpar

\begin{document}
\title{Quasar Radio Dichotomy: Two Peaks, or not Two Peaks, that is the Question}
\author{\v{Z}eljko Ivezi\'{c}, Gordon T. Richards, Pat B. Hall, Robert H. Lupton, Anjoli S. Jagoda, 
Gillian R. Knapp, James E. Gunn, Michael A. Strauss, David Schlegel, William Steinhardt,
Robert J. Siverd}
\affil{Dept. of Astroph. Sciences, Princeton University, Princeton, NJ 08544}

\begin{abstract}
Recent claims by Ivezi\'{c} et al. (2002) that the distribution of 
the radio-to-optical flux ratio, $R$, for quasars is bimodal (the so-called 
quasar radio dichotomy) were questioned on statistical grounds by 
Cirasuolo et al. (2003). We apply the approach suggested by 
Cirasuolo et al. to a sample of $\sim10,000$ objects detected by 
SDSS and FIRST, and find support for the quasar 
radio dichotomy. The discrepancy between the claims by Cirasuolo et 
al. and the results presented here is most likely because
1) the $\sim$100 times larger sample based on two homogeneous surveys
that is used here allows a direct determination of the $R$ distribution, 
rather than relying on indirect inferences based on Monte Carlo 
simulations of several heterogeneous surveys 2) the accurate SDSS 
colors and redshift information allow robust determination of the
K-correction for $R$, which, if unaccounted for, introduces significant
scatter that masks the intrinsic properties of the quasar $R$
distribution.
\end{abstract}

\phantom{x}\vskip -0.5in \phantom{x}

\section{What Statistics to Use?}

\begin{figure}
\plotfiddle{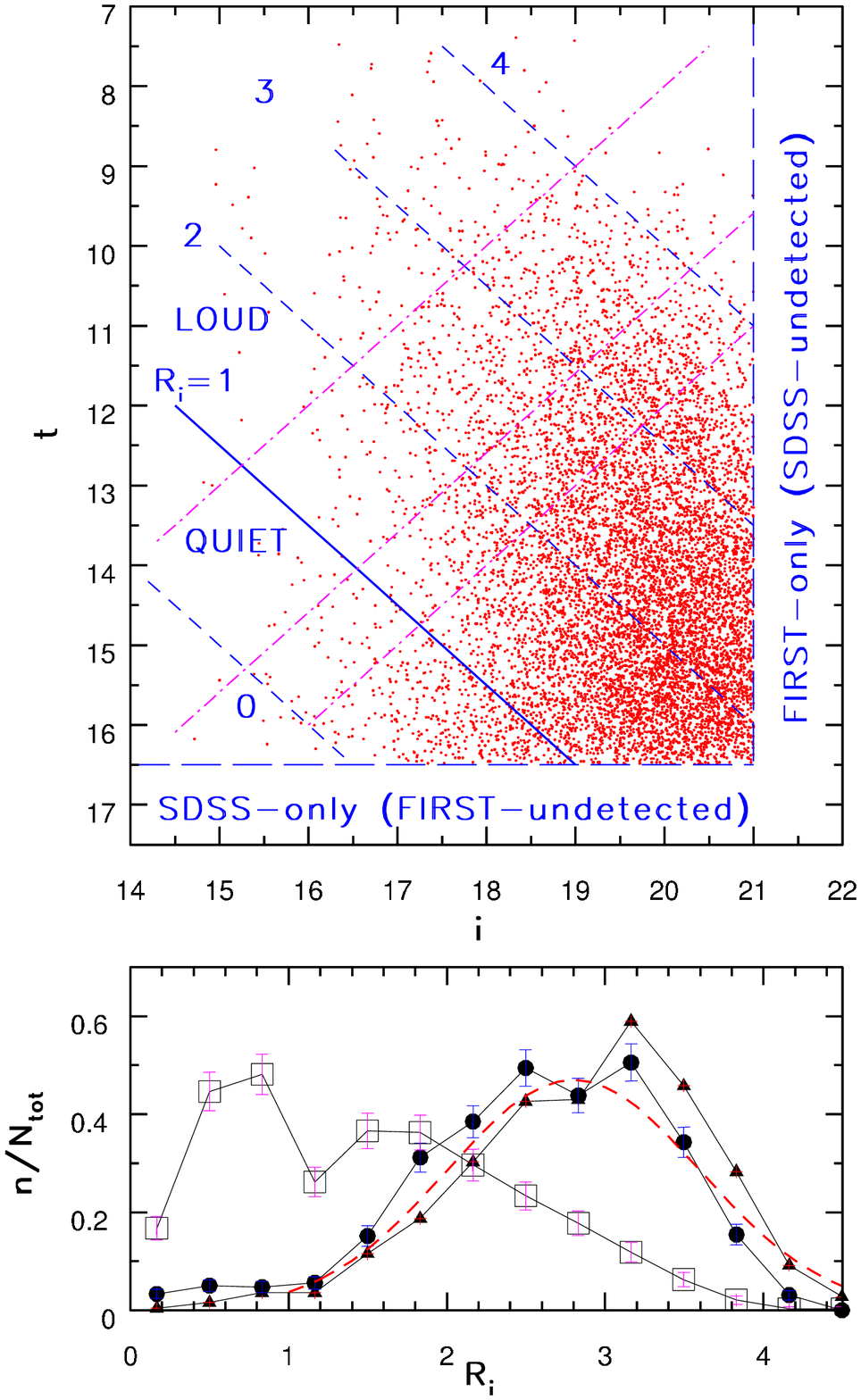}{4.5cm}{0}{40}{40}{-210}{-160}
\plotfiddle{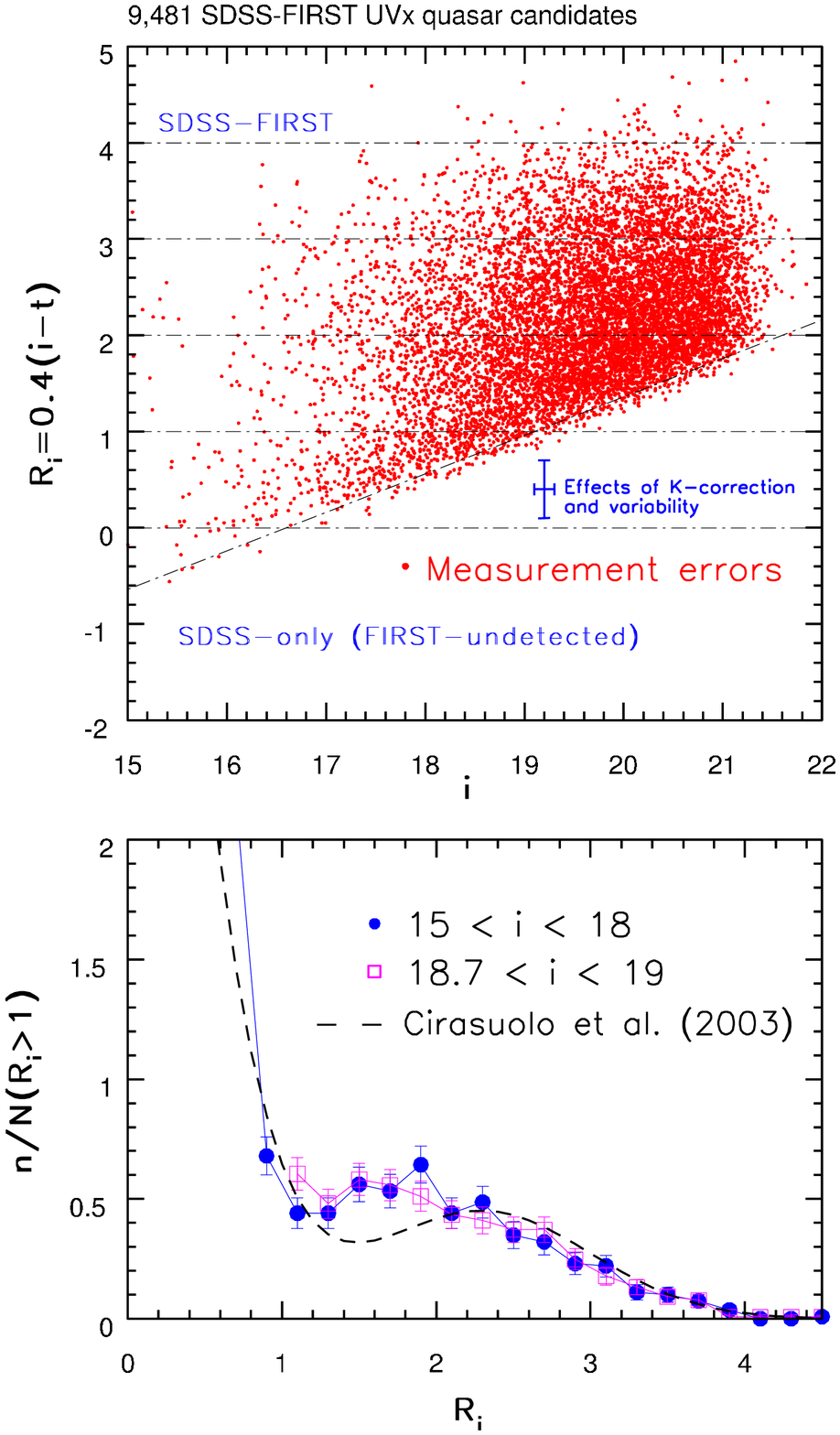}{4.5cm}{0}{40}{40}{-40}{-20}
\caption{The two left panels summarize the analysis of the quasar radio 
dichotomy by Ivezi\'{c} et al. (2002), and are repeated here with a $\sim$3 
times larger sample from SDSS and FIRST ($\sim$10,000 objects). In the top left
panel, which shows the source distribution in the $t$ (radio AB magnitude)
vs. $i$ (optical magnitude) plane, the diagonal dot-dashed lines define regions 
that were used to determine the $R_i=0.4\,(i-t)$ distribution. The $R_i$ 
histograms for these regions, marked by filled circles and triangles in the 
bottom left panel, were interpreted as evidence for a quasar radio dichotomy.
The histogram marked by open squares shows the $R_i$ distribution for 
sources with $i<$ 18, and is shown as an example of a biased estimate
of the $R_i$ distribution. The upper right panel shows the $R_i$ vs. $i$ distribution 
for the same SDSS-FIRST data set as in the two left panels (note that this diagram 
is a sheared, and not simply a rotated, version of the diagram in the top left panel).
The large dot in the top right panel illustrates the typical measurement
uncertainty. The error bars show the uncertainty in $R_i$ 
($\sim$0.2-0.3) mostly due to optical and radio K-corrections,
and in $i$ ($\sim$0.1 mag), due to optical variability.
The two histograms in the bottom right panel (symbols with error bars)
show $p(R_i|i)$ for two ranges of $i$, as marked. The dashed line in the bottom 
right panel shows a best-fit result for $p(R_i|i)$ by Cirasuolo et al. (2003), 
displayed here for illustration (it is shifted left by 0.4 mag to account for 
different optical bands, $i$ vs. $B$). 
}
\end{figure}

There is controversy in the literature about the existence of a bimodality 
in the distribution of radio-to-optical flux ratio, $R$, for quasars (the 
so-called quasar radio dichotomy). For example, White {\em et al.} (2000) 
suggested that previous detections of radio dichotomy were caused by
selection effects. On the other hand, Ivezi\'{c} et al. (2002, hereafter I02)
claimed that a sample of quasars detected by the SDSS and FIRST surveys supports
the existence of a radio dichotomy. The latter result was recently questioned 
on statistical grounds by Cirasuolo et al. (2003, hereafter C03). I02 determined 
the distribution of $R_i=0.4\,(i-t)$ for narrow regions in the $t$ (radio AB magnitude) 
vs. $i$ (optical magnitude) plane that were oriented perpendicular to the $R_i$=const. 
lines (see top left panel in Figure 1). In other words, the quasar density in
the $t$ vs. $i$ plane, $\rho(i,t)$, was found to be a separable function 
$\rho(i,t) = f(R_i) \, g(i+t)$. The $R_i$ distribution, $f(R_i)$, determined this way 
has a strong maximum at $R_i\sim2$, and declines towards smaller $R_i$ (bottom left
panel in Fig. 1). Since a large majority ($\sim90\%$) of quasars undetected by 
FIRST form another peak at $R_i < 0$, the local minimum at $R_i\sim$ 0--1 implies 
the existence of a radio-dichotomy.
 
C03 claimed that a more meaningful quantity is the conditional probability 
distribution $p(R_i|i)$, that is, the $R_i$ distribution for a given 
(narrow range of) $i$, with $\rho(i,t)=p(R_i|i)\,n(i)$. Here $n(i)$ is the 
differential $i$ distribution (``optical counts''). For comparison with their
work, in this contribution we analyze 
the behavior of $p(R_i|i)$. In the top right panel in Figure 1, we 
show the $R_i$ vs. $i$ distribution for $\sim$10,000 quasar candidates detected by 
both SDSS and FIRST (for more details see York 2000, I02, Schneider et al. 2003, and 
references therein). The corresponding $p(R_i|i)$ displayed in the bottom right panel
does not decrease smoothly with $R_i$; 
rather, it suggests a possible local minimum around $R_i\sim1.2$, and a local maximum
around $R_i\sim1.8$. This distribution is consistent with the C03 best-fit shown
by the dashed line in the lower right panel (the latter is in fact a bimodal function).
Note that, given the FIRST flux limit shown as the diagonal 
dot-dashed line in the top right panel, only quasars {\it brighter} than $i\sim19$ 
can be used to directly constrain the position of the local minimum 
in $p(R_i|i)$, and thus a large area optical survey such as SDSS is required
(as opposed to a deeper survey of a smaller area).

\vskip -0.35in
\phantom{x}

\section{To K-correct, or not? }
\vskip -0.25in
\phantom{x}

When analyzing the $R_i$ distribution, it is important to realize
that the scatter due to K-corrections and quasar variability is
much larger than the measurement errors. The uncertainty in $R_i$ 
($\sim$0.2-0.3) is mostly due to optical and radio flux K-corrections,
and optical variability. Even if the intrinsic $R_i$ were the same for 
all quasars (i.e. a $\delta$-function), its observed distribution would 
still have a finite width because of this uncertainty.
In practice, this effect smears any features in $R_i$ distribution
and would reduce any bimodality, if not taken into account.

The need to account for the K-correction can be inferred from the 
improved agreement between different $R_i$ histograms when the 
sample is divided into redshift bins. We compared the $p(R_i|i)$ distributions 
in different redshift bins using the {\it uncorrected} $R_i$, and found them
to be systematically different. Furthermore, the differences between the
$p(R_i|i)$ distributions for different $i$ bins in a given narrow redshift bin 
are smaller than when the whole redshift range is considered. This systematic 
behavior disappears when a proper K-correction for $R_i$ is applied.

We determined the K-correction for $R_i$, such that $R_i^{corr}=R_i^{obs}+\Delta R_i$, as 
\begin{equation} 
\Delta R_i = (\alpha_{radio}-\alpha_{optical})\,\log(1+z),
\end{equation} 
where $\alpha_{radio}$ and $\alpha_{optical}$ are radio and
optical spectral slopes, respectively ($F_\nu \propto \nu^\alpha$).
We use the difference between $g-i$ color for a particular source 
and the median $g-i$ color at the redshift of that source to estimate 
the optical spectral slope (Richards et al. 2003). For radio spectral
slope we assume $\alpha_{radio}$ = -0.5, which is the median value
of radio spectral index for a sample of $\sim$400 quasars with SDSS, GB6, 
FIRST, NVSS and WENSS detections (Ivezi\'{c} et al., in prep.). 

\begin{figure}[t]
\plotfiddle{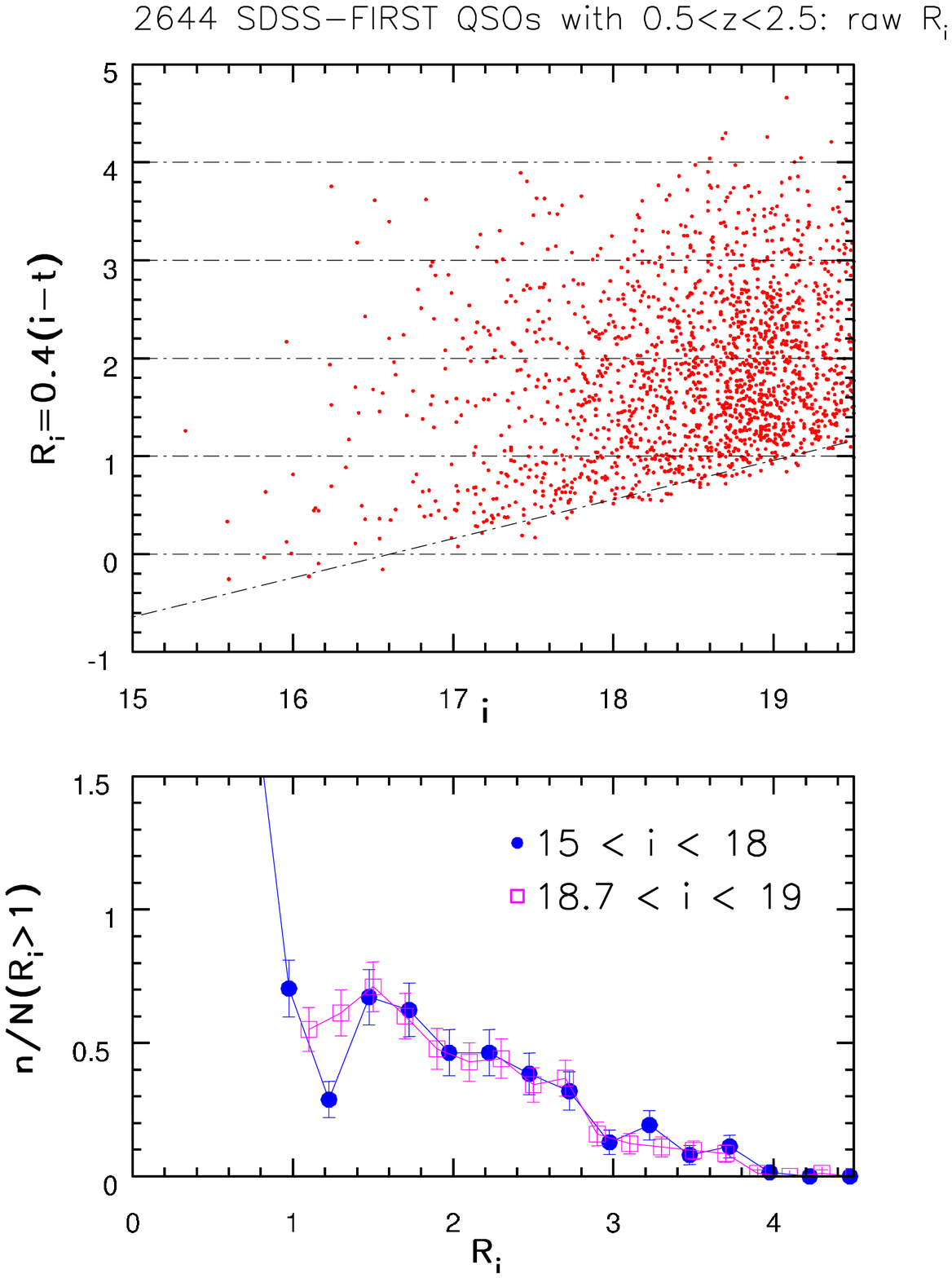}{3.4cm}{0}{40}{40}{-210}{-185}
\plotfiddle{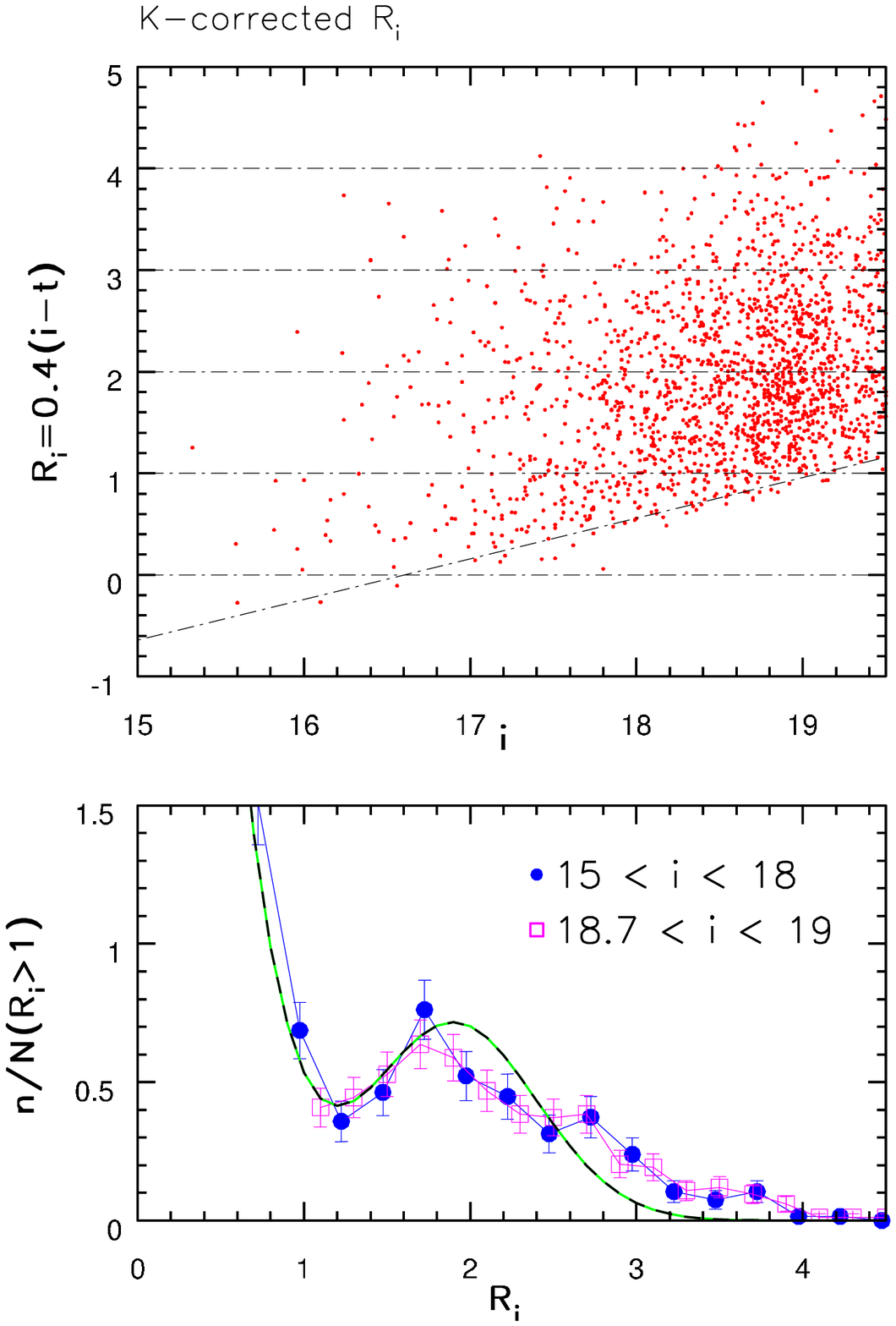}{3.4cm}{0}{40}{40}{-40}{-72}
\caption{
The two left panels show the $R_i$ vs. $i$ and $p(R_i|i)$ distributions for quasars 
with $0.5 < z < 2.5$ in two $i$ magnitude bins, where $R_i$ is {\bf not} 
K-corrected. The two right panels show analogous results when $R_i$  
{\bf is} K-corrected using eq.~1. The thick line 
in the bottom right panel is the best-fit using the same functional form 
as proposed by C03 (a double Gaussian). It has a local minimum at 
$R_i\sim1.2$ and a local maximum at $R_i\sim1.9$, with the maximum-to-minimum 
ratio of $\sim$2.
}
\end{figure}
\vskip -0.35in
\phantom{x}

\section{   Evidence for Quasar Radio Dichotomy }
\vskip -0.25in
\phantom{x}

Figure 2 compares the distribution of SDSS-FIRST quasars with redshifts
in the range 0.5--2.5 in the $R_i$ vs. $i$ plane when $R_i$ is {\bf not} 
K-corrected (left), and when $R_i$  {\bf is} K-corrected using eq.~1 (right).
As evident in the bottom panels, accounting for K-correction increases the significance 
of the detected bimodality. It is important to use
an estimate of the optical spectral slope on an {\it object-by-object}
basis -- it is insufficient to use a mean slope as obtained from 
e.g. a composite quasar spectrum.

\phantom{x}
\vskip -0.55in
\phantom{x}

The dashed line in the bottom right panel in Figure 2 
is the best-fit using the same functional form proposed by C03 (a double 
Gaussian). It has a local minimum at $R_i\sim1.2$ and a local 
maximum at $R_i\sim1.9$, with the maximum-to-minimum ratio of 
$\sim$2. As reported by Ivezi\'{c} et al. (2002), the fraction of 
sources with $R_i > 1$ is 8$\pm$1 \%. The remaining 92\% of quasars,
most of which are not detected by FIRST, are responsible for 
the steep rise of $p(R_i|i)$ for $R_i < 1$. 

We conclude that accurate optical and radio measurements for
a large and homogeneous sample of radio quasars obtained by SDSS 
and FIRST provide conclusive evidence for the existence of
the quasar radio-dichotomy.

\vskip -0.15in
\phantom{x}

{\small 
Funding for the creation and distribution of the SDSS Archive has been provided by 
the Alfred P. Sloan Foundation, the Participating Institutions, the National Aeronautics 
and Space Administration, the National Science Foundation, the U.S. Department of Energy, 
the Japanese Monbukagakusho, and the Max Planck Society. The SDSS Web site is 
http://www.sdss.org/. 
}
\vskip -0.4in
\phantom{x}


\begin{references}
\vskip -0.3in
\phantom{x}
\reference Cirasuolo, M., et al. 2003, astro-ph/0306415
\reference Ivezi\'{c}, \v{Z}, et al. 2002, AJ, 124, 2364
\reference Richards, G.T., et al. 2003, AJ, 126, 1131
\reference Schneider, D.P., et al. 2003, AJ, in press, astro-ph/0308443
\reference White, R.L., et al. 2000, ApJS, 126, 133
\reference York, D.G., et al. 2000, AJ, 120, 1579
\end{references}
\end{document}